\documentclass[floatfix,11pt,
amsmath,amssymb]{revtex4}

\usepackage{epsf}
\usepackage{graphicx}  
\usepackage{dcolumn}   
\usepackage{bm}        
\usepackage{color}
\usepackage{hyperref}
\newcommand{\be}{\begin{equation}}
\newcommand{\en}{\end{equation}}
\newcommand{\bea}{\begin{eqnarray}}
\newcommand{\ena}{\end{eqnarray}}
\newcommand{\sch}{Schwarzschild}

\begin{document}

	\title{Black Holes and Non-perturbative Gravitational Waves in $f(R)$ Gravity }
	\author{Chunmei Liu$^{1}$ and Hongsheng Zhang$^{1}$\footnote{Electronic address: sps\_zhanghs@ujn.edu.cn} }
	\affiliation{$^1$ School of Physics and Technology, University of Jinan, 336 West Road of Nan Xinzhuang, Jinan, Shandong 250022, China}
	
\begin{abstract}
	
		 Exact solutions of spherically symmetric black hole  and gravitational wave are explored in $f(R)$ gravity in arbitrary dimension. We find two exact solutions for the radiation and absorption of null dust. In the framework of general relativity, the Birkhoff theorem strictly forbids the existence of spherical gravitational waves in vacuum space. We find spherical non-perturbative gravitational waves, which are shear-free, twist-free, but expanding.

	\end{abstract}
\maketitle


	\section{Introduction}
     An extensively discussed gravity theory in the literature is $f(R)$-gravity \cite{Sotiriou:2008rp, Capozziello:2011et, Vollick:2003aw, Nojiri:2007uq, Shojai:2011yq, Buchdahl:1970ynr, Starobinsky:2007hu}. Theoretical and observational motivations suggest that Einstein's gravity should be extended to or corrected at UV/IR limits. The theoretical reason is that general relativity is nonrenormalizable. In 1977, Stelle showed that quadratic curvature gravitational actions are renormalizable \cite{Stelle:1976gc}. This discovery was followed by a surge of interest, that was boosted again later on by the discovery of the potential cosmological consequences of these theories, as found by Starobinsky and others \cite{Starobinsky:1980te}. There is also a motivation for $f(R)$ gravity that the theory avoids Ostrogradsky instability \cite{Woodard:2015zca}. The current observations such as Type Ia Supernovae \cite{SupernovaSearchTeam:1998fmf, SupernovaCosmologyProject:1998vns}, and cosmic microwave background (CMB) radiation \cite{Planck:2015fie}, revealed that the universe is undergoing  accelerated expansion. $f(R)$-gravity is proposed to explain the accelerated expansion of the universe \cite{Myrzakulov:2012qp, Sotiriou:2008ve, Nojiri:2007as, Nojiri:2007cq,    Cognola:2007zu, Nojiri:2010wj}. 

	Black holes (BHs) are believed to play a central role in many astrophysical processes, ranging from the evolution of stars and galaxies to powering
	Active Galactic Nuclei and extreme bursts of gravitational radiation in binary BH coalescence \cite{Herdeiro:2014goa}. The  first BH image \cite{EventHorizonTelescope:2019dse} has consolidated the evidence for their physical reality. Central to our understanding of BH physics is the uniqueness theorems \cite{Fabian:2005hr}, stating that the only stationary, regular, asymptotically flat BH solution of the vacuum Einstein gravity
	is the Kerr metric. The uniqueness theorem becomes invalid in modified gravity. Non-\sch~black hole in spherical symmetry~and non-Kerr black hole in rotating symmetry are permitted in $f(R)$-gravity. Some BH solutions in $f(R)$ gravity are studied in detail in the ref \cite{Zhang:2021sjx, Guo:2013swa, Huang:2022urr, Zhang:2014kla, Sotiriou:2011dz, Zhang:2014ala, Zhang:2014goa}. A great deal of attention has also been paid to the issue of spherically symmetric solutions of the gravitational field equations in $f(R)$ gravity \cite{Capozziello:2007wc, Hollenstein:2008hp, Multamaki:2006zb}. There exists a second branch of static, spherically symmetric black holes, over and above the Schwarzschild solutions in models of the $f(R)$ gravity \cite{Multamaki:2006zb, Sebastiani:2010kv,   Capozziello:2007id, Carames:2009ek}. In Ref.\cite{Clifton:2005aj} ,the authors present a class of exact spherically symmetric solutions for the specific choice $f(R)=R^{\delta
		+1}$.
  
   The LIGO and Virgo Collaborations successfully discover gravitational waves from a binary black hole merger \cite{LIGOScientific:2016lio}. The discovery of gravitational waves has opened up a new window to explore the Universe. The proposed space-based GW observatories Taiji, TianQin, and LISA detect GWs in the low-frequency regimes $  10^{-4}-10^{-1} $ Hz \cite{Gong:2021gvw, Jiang:2023hrw}. These facilities are expected to detect more and more gravitational wave events in the future. The exact solution of the first gravitational wave in history was discovered by Brinkmann in 1925 \cite{Brinkmann:1925fr}. The subsequent rigorous solutions of gravitational waves attract a considerable amount of attention \cite{Griffiths:2009dfa}.
   
  Vaidya finds an important model for a
  spherically symmetric spacetime in general relativity \cite{Lindquist:1965zz},
  \be
  ds^{2}=-(1-\dfrac{2M(u)}{r})du^2+2 du dr+r^{2}d\Omega_2^2.
  \label{metricv}
  \en
  In \cite{Zhang:2014kla} Vaidya-like solutions in three or four dimensions are found in \cite{Zhang:2021sjx}. We aim to derive Vaidya-like solutions in the $n$-dimensional $f(R)$ gravity. In some special cases, these BH solutions yield gravitational wave solutions. Then we study the properties of $n$-dimensional gravitational waves. We find that gravitational waves are shear-free, and twist-free but expanding.

  The paper is organized as follows. In Section \uppercase\expandafter{\romannumeral2}, the action of
  $f(R)$ gravity is introduced, and the gravitational field equations are presented \cite{Hu:2016hpm, Zhang:2015nwy, Abdusattar:2021wfv, Guo:2013fda, Cai:2001dz,Zhang:2014kla}. In Section \uppercase\expandafter{\romannumeral3}, we obtain the black hole solution in Vaidya-like coordinates in $f(R)$ gravity. In Section \uppercase\expandafter{\romannumeral4}, we obtain Vaidya-like black holes and gravitational waves  in arbitrary dimensions. In Section  \uppercase\expandafter{\romannumeral5}, we discuss the properties of gravitational waves. In Section \uppercase\expandafter{\romannumeral6}, we summarise our results.

 \section{Field equation in n-dimensional \texorpdfstring{$f(R)$} . spacetime}
   
 The action of the gravitational field in the $f(R)$ gravity model is,
 \be
 S=\frac{1}{16\pi G }\left(\int d^ {n}x\sqrt{-{\rm det}(g)}~f(R)\right)+S_{m}, 
  \label{action}
 \en
  where $G$ is the Newtonian constant, $n$ is the dimension, $g$ is the metric tensor, $R$ is Ricci scalar, and $f(R)$ is an arbitrary function of the Ricci scalar $R$  and $S_m$ denotes matter action. The corresponding field equations reads,
\be
  f_{R}R_{\mu \nu }-\frac{1}{2}fg_{\mu \nu }-\nabla _{\mu }\nabla
 _{\nu }f_{R}+g_{\mu \nu }\square f_{R}=8\pi G T_{\mu \nu },
 \label{field1}
 \en
 where$f_R=\frac{\partial f}{\partial R}$, and $T_{\mu\nu}$ presents the stress-energy for matter fields. For $f(R)=R^{d+1}$, the field equation becomes,
\be
 \left( d+1\right) R^{d}R_{\mu \nu}-\frac{1}{2}R^{d+1}g_{\mu \nu }-(d+1)\nabla _{\mu }\nabla
 _{\nu }R^{d}+g_{\mu \nu }\left( d+1\right) \square R^{d}=8\pi G T_{\mu\nu}.
 \label{field2}
 \en
	  
   We investigate weak-field limit of $f(R)$ gravity. The Poisson equation in $f(R)$ gravity reads \cite{Clifton:2005aj},
   \be
   \nabla ^2\Phi \sim \rho R^{-d}.
   \en
     The Poisson's equation for Newtonian gravitational potential is $\nabla ^2\Phi \sim \rho$. Compared to the Poisson's equation,  $f(R)$ gravity leads to an extra factor of $R^{-d}$. This can be interpreted as a variable Newtonian constant. This is reasonable since scalar-tensor theory leads to a variable Newtonian constant, and $f(R)$ gravity is conformally equivalent to scalar tensor theory.

 \section{Black hole solutions in arbitrary dimensions}	
 In the following text, we set $8\pi G = 1$. The limit $d\rightarrow 0$ reduces Einstein-Hilbert Lagrangian. 
 The black hole solutions have been found in the following coordinates \cite{Zhang:2014kla},
 \be
 ds^2=-A(r)dt^2+\frac{1}{B(r)}dr^2+r^2d\Omega_k^2,
 \label{metric}
 \en
 where the metric fields $A$ and $B$ are both  functions of $r$. Here, $d\Omega_k^2=h_{ij}(x)dx^{i}dx^{j}$ is the $n-2$-dimensional metric  with 
 constant curvature.
 
 In this section, we re-explore the BH
 solutions in the n-dimensional $f(R)$ gravity in Vaidya-like coordinates. 
 The Vaidya solution \cite{Vaidya:1951zza} is important since it encodes some essential properties of the
 dynamical spherically symmetric spacetimes while
 remaining simple enough to handle. In the n-dimensional spacetime with a maximally symmetric (n-2)-subspace, exact generalized Vaidya and Vaidya-like solutions are considered in the context of $f(R)$ gravity. We transform the coordinate (\ref{metric})$(t,r,x^i)$ to $(u,r,x^i)$, where $u=t-r_{*}$, $r_{*}=\int\dfrac{dr}{\sqrt{AB}}$ \cite{Culetu:2016ewn}.
 The metric ansatz reads,
\be
 ds^{2}=-A(r)du^2-2\sqrt{{A(r)}/{B(r)}} du dr+r^{2}h _{ij}dx^{i}dx^{j},
  \label{metrics1}
 \en
 where $r^2h_{ij}dx^idx^j$ stand for the metrics of $n-2$-dimensional manifold
 with constant curvature.

The expression for $h_{ij}dx^idx^j=d\Omega_k^2$ is given by,
\begin{equation}
	d\Omega_k^2=\left\{
	\begin{array}{lr}
		d\theta_1^2+\sum\limits_{i=2}^{n-2}\prod\limits_{j=1}^{i-1}\sin^2 \theta_j d\theta_i^2 & k=1~, \\
		d\theta_1^2+\sum\limits_{i=2}^{n-2}\prod\limits_{j=1}^{i-1}\sinh^2 \theta_j d\theta_i^2 & k=-1~, \\
		\sum\limits_{i=1}^{n-2}d\theta_i^2 & k=0~. \\
	\end{array}
	\right.
\end{equation}
  There are three types of horizons, elliptic horizons with $k = 1$, flat horizons with $k = 0$, and hyperbolic horizons with $k =-1$.
  The metric function $h_{ij}$ is a function of the coordinates $x^{i}$ only, and we shall refer to this metric as the horizon
metric. The horizon is denoted by $M^{n-2}$. 

From the action and considering vacuum, the
equations of motion are, 
\begin{equation}
	\begin{aligned}
		\chi_{\mu \nu}&\equiv \left( d+1\right) R^{d}R_{\mu \nu}-\frac{1}{2}R^{d+1}g_{\mu \nu }-(d+1)\nabla _{\mu }\nabla
		_{\nu }R^{d}+g_{\mu \nu }\left( d+1\right) \square R^{d}\\
		&=0,
		\label{field3}
	\end{aligned}
\end{equation}
and the corresponding components of $\chi_{u u}$  and $\chi_{i j}$are
\begin{equation}
\begin{aligned}
	&(d+1) \left(A' \left(-\frac{B A'}{2 A}+\frac{B'}{2}+\frac{(n-2) B}{r}\right)+B A''\right) \\
	&-\frac{d (d+1) A \left(R'(r) \left(r B'+2 (n-2) B\right)+2 r B R''(r)\right)}{r R(r)}\\
	&-\frac{2 d \left(d^2-1\right) A B R'(r)^2}{R(r)^2}+A R(r)=0,
\end{aligned}
	\label{eq:tt}
\end{equation}

\begin{equation}
	\begin{aligned}
	&\frac{(d+1) r B A' \left(\frac{d r R'(r)}{R(r)}-1\right)}{A}+\frac{d (d+1) r \left(R'(r) \left(r B'+2 (n-3) B\right)+2 r B R''(r)\right)}{R(r)}\\
	&-(d+1) r B'+\frac{2 d \left(d^2-1\right) r^2 B R'(r)^2}{R(r)^2}+2 (d+1) (n-3) (k-B)-r^2 R(r)=0.
	\end{aligned}
\end{equation}

  Here a prime denotes derivative with respect r, where $ R_{ij}(h)$ is the Ricci tensor of the horizon metric, 
\begin{equation}
	R_{ij}(h)= k(n-3)h_{ij}.
\end{equation}
  This makes it possible to construct black hole solutions with a non-spherical horizon topology.
 
The definition of the curvature scalar, provided from the metric, is given by
 \begin{align}
 	R&=-\frac{2 B \left((n-2) A'+r A''\right)+r A' B'}{2 r A}+\frac{B A'^2}{2 A^2} \nonumber \\
 	&+\frac{-(n-2) r B'-(n-3) (n-2) B+{{\cal R}(h)}}{r^2},
 	\label{Riccis}
 \end{align}
  where ${\cal R}(h)$ is $n-2$-dimensional Ricci scalar, ${\cal R}(h)= k(n-2)(n-3)$. For the moment, the non-vanishing Riemann tensor components are easy to obtain. 

The Riemann tensor of $n-2$-dimensional maximum symmetric space is given by,
\be
 {\cal R}(h)_{ijkl}=k(h_{ik}h_{jl}-h_{il}h_{jk}).
\en

From the above equation,  we can obtain
the generalized BH solution in the n-dimensional $f(R)$ gravity. For this case, the metric coefficients A and B from the relations  (\ref{field2}) and (\ref{Riccis}) are obtained as,

\be
A(r)=r^{-\frac{4 d (2 d+1)}{2 d-n+2}} \left(C r^{\frac{8 d^2-4 d (n-3)+n^2-5 n+6}{2 d-n+2}}+k\right),
\en
\be
 B(r)=-\frac{\left((n-3) (-2 d+n-2)^2\right) \left(C r^{\frac{8 d^2-4 d (n-3)+n^2-5 n+6}{2 d-n+2}}+k\right)}{\left(4 d^2+4 d-n+2\right) \left(8 d^2-4 d (n-3)+n^2-5 n+6\right)},
 \en
where $C$ is an integration constant. The situations of
different $k$ are in analogy to the three different spatial
geometries at r = constant in the case of the topological
Schwarzschild solution. In $f(R)$ gravity, constant $C$ is related to the mass of black hole. The solutions reduce to the Schwarzschild solutions in the limit $d\rightarrow 0$ and $n\rightarrow 4$, where $C=-2m$, and $m$ is the ADM mass of the black hole. When $d\neq 0$, determining $C$ is a complex problem. The constant is related to the Misner-Sharp mass through,     
\be
2m=(-C)^{(1-d)/(1-2d+4d^2)},
\label{ Misner-Sharp mass}
\en
where $m$ is the Misner-Sharp mass, for detailed demonstration of this equation see \cite{Zhang:2014goa}.
The Ricci scalar vanishes when $k=0$, while the Kretschmann scalar does not vanish. The result for the Kretschmann scalar\cite{Henry:1999rm} is 
\be
	\begin{split}
	R_{abcd}R^{abcd}&=\frac{(n-3)^2 (n-2) (n-1) (-2 d+n-2)^2}{(-4 d (d+1)+n-2)^2 \left(8 d^2-4 d (n-3)+n^2-5 n+6\right)^2}C^2 r^{\frac{4 (2 d+1) d}{2 d-n+2}+4 d-2 n+2}\\&(32 d^3 (8-3 n)+4 d^2 (n (13 n-63)+76)+64 d^4-12 d (n-3) (n-2)^2+(n-3) (n-2)^3).
\end{split}
\en
Thus it still describes a curved spacetime. 
 The apparent horizon $l(r)$ is defined as the trapped surface, which satisfies,
\be
\gamma^{ab}\partial_{a}l\partial_{b}l=0,
\en
where $\gamma$ metric is given by,
\be
\gamma=-A(r)dt^2+\frac{1}{B(r)}dr^2.
\en
The apparent horizon dwells at
\be
r_{a}=\left(-\frac{k}{C}\right)^{\frac{2 d-n+2}{8 d^2-4 d n+12 d+n^2-5 n+6}}.
\en

 \section{Vaidya-like black holes and spherical gravitational waves in arbitrary dimensions}
 
 Inspired by the Vaidya metric, we consider a metric ansatz with the form, 
\be
 ds^{2}=-A(u,r)du^2\pm2\sqrt{{A(u,r)}/{B(u,r)}} du dr+r^{2}h _{ij}dx^{i}dx^{j}.
 \label{metrics2}
 \en 
 We write $A$ and $B$ as follows, 
 \be
 A(u, r){=}r^{\frac{2 d (2 d+1)}{-d+\frac{n}{2}-1}} \left(\mathcal{G}(u) r^{\frac{8 d^2-4 d (n-3)+n^2-5 n+6}{2 d-n+2}}+k\right)
 \label{solution1}
 \en
 and
 \be
 B(u, r){=}-\frac{\left((n-3) (-2 d+n-2)^2\right) \left(\mathcal{G}(u) r^{\frac{8 d^2-4 d (n-3)+n^2-5 n+6}{2 d-n+2}}+k\right)}{\left(4 d^2+4 d-n+2\right) \left(8 d^2-4 d (n-3)+n^2-5 n+6\right)}.
 \label{solution2}
 \en
 The function  $\mathcal{G}(u)$  is arbitrary smooth functions of 
 $u$ in the second order, where the Ricci scalar $R$ is given by
 \be
 R=\frac{4 d (d+1) k (n-3) (n-1)}{r^2 (4 d (d+1)-n+2)}.
 \en
  Substituting $A$ and $B$ into the field equation, one obtains,
 \be
T_{ab}=\Phi(u,r)(k)_{a}(k)_{b},
\label{source1}
\en
where,
 \be
 k_{a}=-(du)_{a} ,\qquad
 k^{a}=\sqrt{\frac{B}{A}}\left(\frac{\partial}{\partial r}\right)^a.
 \en
 $k_{a}$ satisfies $k_{a}k^{a}=0$. A null matter propagating along $k^{a}$ is described by (\ref{source1}). And the function $\Phi(u,r)$ reads,
 \be
 \Phi(u,r)=\mp\frac{ (d+1) (2 d-n+2)  r^{2 d-n+2} \left(R\right)^d}{2\sqrt{-\frac{(4 d (d+1)-n+2) \left(8 d^2-4 d (n-3)+n^2-5 n+6\right) }{(n-3) (-2 d+n-2)^2}}}\frac{d\mathcal{G}}{du}.
 \label{source}
 \en 
 
One sees that the cases $k=\pm1$ describes dynamical spacetime with sources(\ref{source}, \ref{source1}). Both the cases of $\pm$ in the metric (\ref{metrics2}) solve the field equation. Lindquist et al \cite{Lindquist:1965zz} find that $-\frac{dm}{du}$ is exactly the total power output for the Vaidya metric. Eq.(\ref{source1}) belongs to the following class of stress energy,
\be
T_{a b}=\varphi^{2}(k)_{a}(k)_{b},
\label{source2}
\en
where $\varphi^{2}$ is a positive define function on the manifold. The stress energy of a massless scalar field can be written in this form (\ref{source2}). So when $\Phi(u,r)$ with 
the ``-" sign, the equation (\ref{source2}) requires $\frac{d\mathcal{G}}{du}<0$. When the case of $\Phi(u,r)$ is taken as ``-" and $\frac{d\mathcal{G}}{du}<0$, it describes a shining star whose mass is decreasing. While  ``+" and $\frac{d\mathcal{G}}{du}>0$ for an absorbing star in $n$-dimensional spacetime \cite{Lindquist:1965zz,Vaidya:1951zza}.  
 
The situation where $k=0$ is particularly intriguing, as it pertains to a purely spherically symmetric gravitational wave. It is clear that gravitational waves propagate in $n$-dimensional spacetime. Both the Ricci scalar and $T_{\mu\nu}$ are  zero when k=0. We obtain a non-perturbative vacuum gravitational wave solution in $R^{d+1}$ gravity without general relativity limit. The solution proves the existence of $n$-dimensional  $R^{d+1}$ gravitational waves. 
 We  check the Kretschmann scalar,
\be
	\begin{split}
		R_{abcd}R^{abcd} = & \frac{(n-3)^2(n-2)(-2d+n-2)^2 r^{\frac{2(8d^2-4d(n-2)+n^2-3n+2)}{2d-n+2}}}{(4d^2+4d-n+2)^2(8d^2-4d(n-3)+n^2-5n+6)^2} \\
		& \Bigg\{(n-1)\mathcal{G}^2 \bigg[4d^2(13n^2-63n+76) - 32d^3(3n-8) + 64d^4 \\
		& \qquad\qquad\quad- 12d(n-3)(n-2)^2 + (n-3)(n-2)^3\bigg] \\
		& + 4\sqrt{2}d(2d+1)(2d-n+2)\frac{d\mathcal{G}}{du}r^{\frac{(n-2)(4d-n+3)}{2d-n+2}+1}\sqrt{\frac{A(u,r)}{B(u,r)}}\Bigg\},
	\end{split}
\label{K4}
\en
 which shows that the spacetime is not flat.
 
  The solution of GW is important. In fact, we can credibly demonstrate the GW solution in a clearly looking forward way. Now we show it in details.
  
  Inspired by the Brinkmann GW solution, we consider a metric ansatz,
  \be
  ds^{2}=-A(u,r)du^2\pm2\sqrt{{A(u,r)}/{B(u,r)}} du dr+r^{2}h _{ij}dx^{i}dx^{j},
  \en 
  where ${{h}_{ij}}d{{x}^{i}}d{{x}^{j}}=\sum\limits_{i=1}^{n-2}d\theta_i^2$. It describes a space of flat horizons of $n-2$-dimensional submanifold of manifold $M$.
 
 The Ricci scalar provided from the metric (\ref{metrics2}) is given by 
 \be
 \begin{aligned}
 	R(u,r)&=-\frac{B(n^2-5n+6)}{r^2}-\frac{(n-2)BA'}{rA}+\frac{BA'^2}{2A^2}-\frac{(n-2)B'}{r}\\
 	&-\frac{A'B'}{2A}-\frac{BA''}{A}
 	+\sqrt{\frac{A}{B}}\left(-\frac{BA'\dot{A}}{A^3}+ \frac{B'\dot{B}}{AB}+\frac{B\dot{A}'}{A^2}-\frac{\dot{B}'}{A}\right)
 \end{aligned}
 \en
 
 Then, according to the above quantities, we solve field equation (\ref{field3}) without source, 
 \begin{equation}
 	\begin{aligned}
 		\chi_{uu}& = \frac{2A \left( d(d+1)R(u,r) \left( R'(u,r) \left( -rB' - 2(n-2)B \right) - 2rBR''(u,r) \right) - 2d(d^2-1)rBR'(u,r)^2 \right)}{rR(u,r)^2} \\
 		&+ 2AR(u,r) + \frac{2(d+1)A'\dot{A}}{A\sqrt{\frac{A}{B}}}-\frac{(d+1) B \left(2 \dot{A}' \sqrt{\frac{A}{B}}+A'^2\right)}{A}+\frac{d+1}{rBR(u,r)}\left( 2drB \dot{B} R'(u,r) \sqrt{\frac{A}{B}} \right) \\
 		&+\frac{d+1}{rB}\left( 2 B^2 \left((n-2) A'+r A''\right)-2 r B' \dot{B} \sqrt{\frac{A}{B}}+B\left( r A' B'+2 \sqrt{\frac{A}{B}} \left((n-2) \dot{B}+r \dot{B}'\right)\right) \right)=0,
 	\end{aligned}
 \end{equation}
 \begin{equation}
 	\begin{aligned}
 		\chi_{ij}&=-r^2 R(u,r)-(d+1)\left( \frac{r A' B}{A}+(2(n-3)B+rB')\right) +\frac{2 d \left(d^2-1\right) r^2 B R'(u,r)^2}{R(u,r)^2}\\
 		&+\frac{d (d+1) r}{A R(u,r)}\left( R'(u,r) \left(r A' B+A \left(r B'+2 (n-3) B\right)\right)+2 r A B R''(u,r)+2 B \dot{R}(u,r) \sqrt{\frac{A}{B}}\right)=0.
 	\end{aligned}
 \end{equation}
 Here a prime denotes derivative with respect $r$, a dot denotes derivative with respect $u$.
 From the above equation,  we obtain
 vacuum dynamic solutions in the n-dimensional $f(R)$ gravity,
 
 \be
 A(u, r){=}r^{-\frac{(n-2) (4 d-n+3)}{2 d-n+2}} \mathcal{G}(u)
 \label{solution3}
 \en
 and
 \be
 B(u, r){=}-\frac{\left((n-3) (-2 d+n-2)^2\right) \left(\mathcal{G}(u) r^{\frac{8 d^2-4 d (n-3)+n^2-5 n+6}{2 d-n+2}}\right)}{\left(4 d^2+4 d-n+2\right) \left(8 d^2-4 d (n-3)+n^2-5 n+6\right)}.
 \label{solution4}
 \en
 $\mathcal{G}(u)$  is an arbitrary second-order differentiable smooth function.  $A(u, r)$ and $B(u, r)$ describe arbitrary spherically symmetric gravitational wave, including the soliton. For example, when $\mathcal{G}(u)=\sin(u)$, the solution describes ordinary gravitational wave, and when $\mathcal{G}(u)= e^{-u^2}$, the energy is distributed in a finite space, so the solution describes soliton. We do not use any approximation in solving the field equation, so it is a non-perturbative gravitational wave solution. When $k=0$, the GW solution  (\ref{solution1}, \ref{solution2}) is consistent with the GW solution (\ref{solution3}, \ref{solution4}).

\section{Properties of the GW}
 It should be emphasized that  (\ref{solution3}, \ref{solution4}) describes a spherically symmetric gravitational wave. This section provides a detailed analysis of the properties of gravitational waves. We consider a congruence of graviton rays analogous to a congruence of light rays. The out going wave vector is,
\be
K_{a}=-(du)_{a}.
\en
To obtain a detailed description of the null congruence of the graviton ray, we use the formalism method to redefine (\ref{metrics2}),
\be
\tau_{1}=du,\quad \tau_{2}=-[g_{00}du+2g_{01}dr]\frac{1}{2},\quad \tau_{\mu}=rdx^{i},
\en
where the range of values for $\mu$ is from 3 to $n$.
The metric for (\ref{metrics2}) can be expressed as,
\be
g=-\tau_{1}\tau_{2}-\tau_{2}\tau_{1}+\sum_{\mu=3}^{n}\tau_{\mu}^{2}.
\en
The graviton congruence is geodesic congruence,
\be
K^{a}\nabla_{a}K_{b}= 0.
\en
Unlike the pp-wave case, the vector field $K$  does not satisfy Killing's equation in this context,
\be
\nabla K=\nabla_{a}K_{b}=\frac{-1}{rg_{01}} (\sum_{\mu=3}^{n}\tau_{\mu}^{2})+\frac{-{g'}_{00}+2\dot{g}_{01}}{2g_{01}} \tau_1^2.
\en
Here a prime denotes derivative with respect $r$, and a dot for $u$. As the equations become lengthy, we present them directly as components of the metric in (\ref{metrics2}). To study the motion of the congruence, we explore its expansion, shear, and twist on the sectional $n-2$-dimensional volume. We denote the projection of a tensor on the sectional volume using a hat symbol. The expansion, shear, and twist are expressed as,
\be
\hat{\theta}=\hat{g}_{ab}(\nabla^a K^b){\hat{}}=-\frac{(n-2)}{rg_{01}},
\en
\be
\hat{\sigma}_{ab}=(\nabla_{(a} K_{b)}){\hat{}}-\frac{1}{2}\hat{\theta}\hat{g}_{ab}=0,
\en
and,
\be
\hat{\omega}_{ab}=(\nabla_{[a} K{_b]}){\hat{}}=0.
\en
It can be demonstrated straightforwardly that the metric given by (\ref{metrics2}) exhibits no shearing or twisting but expansion, which is in agreement with our physical understanding of radial gravitational waves. The corresponding variation ratios are expressed by $\hat{\theta}$. The Raychaudhuri equation \cite{Raychaudhuri:1953yv} for the affine parameterized null congruence can be employed to describe the behavior, 
\be
K^a\nabla_a \hat{\theta}=-\frac{1}{2}\hat{\theta}^2-\hat{\sigma}_{ab}\hat{\sigma}^{ab}+ \hat{\omega_{ab}}\hat{\omega^{ab}}-R_{ab}K^aK^b=-\frac{1}{2}\frac{(n-2)^2}{r^2g^2_{01}}-\frac{(n-2)g'_{01}}{rg^3_{01}}.
\en
Under the condition $k = 0$, the resulting values for $g_{01}$ and $\hat{\theta}$ are given by,
\be
g_{01}=-\sqrt{-\frac{\left(4 d^2+4 d-n+2\right) \left(8 d^2-4 d (n-3)+n^2-5 n+6\right) r^{-\frac{8 d^2-4 d (n-3)+n^2-5 n+6}{2 d-n+2}-\frac{(n-2) (4 d-n+3)}{2 d-n+2}}}{(n-3) (-2 d+n-2)^2}},
\en
\be
\hat{\theta}=\frac{(n-2) r^{\frac{4 d^2+n-2}{2 d-n+2}}}{\sqrt{-\frac{\left(4 d^2+4 d-n+2\right) \left(8 d^2-4 d (n-3)+n^2-5 n+6\right)}{(n-3) (-2 d+n-2)^2}}}.
\label{expand}
\en
The explicit calculation result of the Raychaudhuri equation is obtain,
\be
K^a\nabla_a \hat{\theta}=-\frac{(n-3) (n-2) \left(8 d^2-2 d (n-4)+(n-2)^2\right) (2 d-n+2) r^{\frac{2 \left(4 d^2+n-2\right)}{2 d-n+2}}}{2 \left(4 d^2+4 d-n+2\right) \left(8 d^2-4 d (n-3)+n^2-5 n+6\right)}.
\label{varition}
\en
Note that $\hat{\theta}$ and $K^a\nabla_a \hat{\theta}$ are not related to $u$. Based on the studies of Vaidya and Kinnersley metrics, one can interpret the parameter $u$ as a time coordinate to some extent. One finds,
\be
\lim_{d\to 0}\hat{\theta}=\frac{n-2}{r},
\en
 and,
\be
\lim_{d\to 0}K^a\nabla_a \hat{\theta}=-\frac{(n-2)^2}{2 r^2},
\en
  which means that they do not vanish even in the limit $d\rightarrow 0$, implying that $d=0$ is a point of discontinuity. This result supports the non-perturbative nature of gravitational waves in the $f(R)$ theory. This discontinuity can be interpreted from the field equation (\ref{field1}). We study $f_R=(1+d)R^{d}$, when $d\neq0$,
  \be
 \square f_{R}\sim R^{d-2}, \qquad \nabla_{\mu } \nabla_{\nu } f_{R}\sim R^{d-2}.
  \en
  These two terms can be very large for a tiny $R$. But when $d=0$, we directly have
  \be
  \square f_{R}=0, \qquad \nabla_{\mu } \nabla_{\nu }f_{R}=0. 
  \en
  This means that these two terms can jump from very large values to zero when $d$ runs from very small value to zero. This is the reason why the discontinuity appears.
  
\section{Conclusion}	
This article is summarized as follows. We analyze vacuum solutions of $f(R)$ theories of gravity in  $d$-dimensional spherically symmetric spacetime. We consider here the modification to the gravitational Lagrangian $R \rightarrow R^{1+d}$. Finding exact solutions in $f(R)$ gravity theory is crucial but challenging because the equations of motion are very complex with higher derivative terms. We find the vaidya-like black hole sourced by null dust in $f(R)$-gravity in arbitrary dimensions. The black hole solutions reduce to the Schwarzschild solution in four-dimensional spacetime in the limit $d\rightarrow 0$.  We obtain exact spherical gravitational wave solutions in $f(R)$-gravity in arbitrary dimensions. We do not use any approximation in solving the field equation, so it is a non-perturbative gravitational wave solution. When $k=0$, the Ricci scalar and $T_{\mu \nu}$ are both zero, (\ref{solution3}, \ref{solution4}) describes a spherically symmetric gravitational wave. We check the Kretschmann scalar show that the space is not Minkowski space. When $k=\pm1$, the solution describes  dynamic spacetimes with sources. The case that $\Phi(u,r)$ is taken as ``-" and $\frac{d\mathcal{G}}{du}<0$  describes a shining star, while  ``+" and $\frac{d\mathcal{G}}{du}>0$ for an absorbing star in $n$-dimensional spacetime.  We investigate the propagating properties of the wave, and find that it is shear-free, twist-free, but expanding.

  \end{document}